# Considerations on the collapse of the wavefunction

J. Reintjes[1] and Mark Bashkansky[2],

[1]Sotera Defense Solutions, Columbia, MD

[2]Code 5613, Naval Research Laboratory, Washington DC, USA 20375-5320

We investigate the specific form that the collapsed quantum state of a signal photon can take when its entangled idler is measured in an entangled ghost imaging configuration using a type II collinear phase matched spontaneous parametric downconversion (SPDC) interaction. Calculation of the correlated counting rate distributions in the ghost image plane and diffraction plane show that agreement between collapse and non-collapse models is obtained if the signal is assumed to collapse into a specific mixed state. However, if the signal is assumed to collapse into a pure state, significant differences arise between the predictions of the two collapse models.

## I. INTRODUCTION

Collapse of the wavefunction plays a central role in the theory of quantum measurement, and has been discussed extensively by many authors. (see, e. g. 1-4 and references therein). Wavefunction collapse has also been discussed in the context of state preparation in which the measurement of one component of an entangled pair projects the remaining (unmeasured) component into an un-entangled single particle quantum state that depends on the result of the measurement. (5-6). There remain conceptual issues with respect to collapse, or reduction, of the wavefunction as to whether it represents a physical process or is just a mathematical concept.

In this paper, we discuss features of the collapse of the wavefunction (or equivalently, for our purposes, projection of the quantum state) within the context of state preparation using entangled signal-idler photon pairs from a spontaneous parametric downconversion (SPDC) source in the entangled ghost imaging configuration. In comparing the two concepts (collapse and non-collapse pictures), we can consider several issues. In the non-collapse model, if the signal is measured in delayed correlation at the diffraction plane, nothing distinguishes the time $t_1$ when the idler is measured. The signal simply evolves continuously to the diffraction plane at time $t_2$ when it is measured. The concept of collapse, and the time of the collapse, do not enter the calculation. However, it has been pointed out that the signal-idler biphoton is a single, entangled quantum system and cannot be considered to be composed of two separate particles with independent properties [7]. Since quantum measurements are non-unitary, and therefore nonlinear, and the optical diffraction equations of the non-collapse model are linear, one might well ask whether the linear evolution of the signal continuously through the time of the non-unitary operation is justified on fundamental grounds. This logic alone might serve to justify the collapse concept as a required part of the physical description of the interaction. However, it is noted in the literature that the collapse and non-collapse models are indistinguishable since they give the same theoretical predictions and cannot be distinguished experimentally. [5,8].

Here we examine the specific form that the quantum state of the signal can take following an idler measurement, and calculate its propagation properties from the ghost image plane to a downstream diffraction plane. We investigate the extent to which the predictions of propagation properties using different collapse models agree with the predictions of non-collapse models. Our results demonstrate that agreement between collapse and non-collapse models can be obtained if the signal is assumed to collapse into a specific mixed state. However, measurable distinctions between collapse and non-collapse models are predicted to arise if the signal is modeled to collapse into a pure state.

## II. SPDC INTERACTION GEOMETRY

The configuration under consideration, shown in Fig. 1, is a modification of the entangled ghost imaging geometry of Pittman. [9]

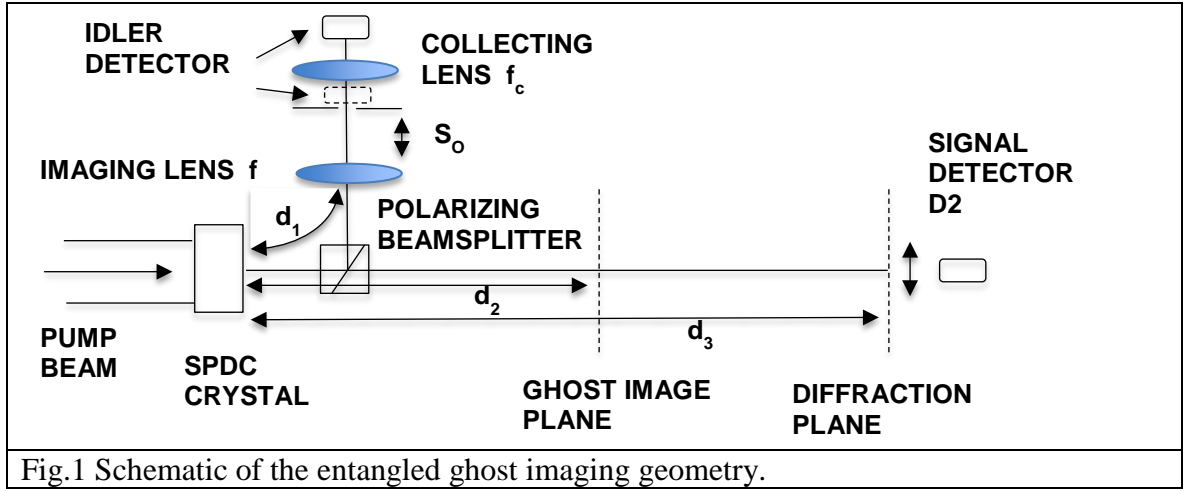

Fig.1 Schematic of the entangled ghost imaging geometry.

A pump beam at a wavelength $\lambda_P$ is incident on a thin BBO crystal cut for collinear type II phase matching. The signal and idler photons are separated with a polarizing beamsplitter. The idler passes through an imaging lens f and then illuminates a slit of width w located at distance $S_o$ after the lens. A collector lens, $f_c$, is placed one focal length behind the slit. A single photon sensitive detector D1 is placed in the idler either at the slit plane or in the focal plane of the collector lens. The signal photon propagates freely through the ghost image plane at a distance $d_2$ from the SPDC crystal to a diffraction plane at a distance $d_3$ from the SPDC crystal. The signal photon is detected with a point detector D2 that is scanned transversely in the detection plane. In our calculations the signal detector is located at either the ghost image plane or the diffraction plane. The position of the ghost image plane, as described in Ref. 9 is defined by the relation

$$\frac{1}{S_0} + \frac{1}{d_1+d_2} = \frac{1}{f} \qquad (1)$$

and the magnification is given by

$$M = \frac{d_1+d_2}{S_0}. \qquad (2)$$

In our analysis we set M = 2 and $S_0 + d_1 = d_2$ so that a measurement of the signal at the ghost image plane is simultaneous with the measurement of the idler at the slit. The measurement of the signal at the diffraction plane is made in delayed correlation with the idler measurement at the slit.

Under the collapse concept, the signal and idler are generated in the SPDC crystal and evolve along their respective paths with linear propagators as an entangled biphoton state. At time $t_1$ the idler is measured, collapsing the biphoton state to a single particle signal state at the ghost plane. This state can then either be measured at the ghost plane at time $t_1$ or it can propagate to the diffraction plane and be measured at a later time $t_2$, where $t_1$ and $t_2$ are related by

$$t_2 - t_1 = d_3/c - d_2/c. \qquad (3)$$

In this approach, the collapsed signal state is measured as a simple expectation value of the signal field operators as there is no longer an idler photon to be correlated with. The counting rate distribution of the signal in the measurement plane is given by the expectation value of $a_s^\dagger a_s$.

In the non-collapse model, the signal and idler fields propagate independently over their respective paths from the SPDC crystal face to their respective detectors at which point the correlated counting rate is measured either in coincidence if the signal is measured at the ghost plane or in delayed correlation if the signal is measured in the diffraction plane. The correlated counting rate in one transverse dimension for the idler detector in plane $z_{mi}$ and the signal detector in plane $z_{ms}$ is given by

$$CCR(x, z_{mi}, x, z_{ms}) = <\psi(z_{mi}, z_{ms})|E_i(x_{mi}, z_{mi})E_s(x_{ms}, z_{ms})E_s^+(x_{ms}, z_{ms})E_i^+(x_{mi}, z_{mi})|\psi(z_{mi}, z_{ms})> \qquad (4)$$

where $E_{s(i)}^+$ is the positive-frequency portion of the signal (idler) electric field operators at the detector and $|\psi(z_i, z_s)>$ is the corresponding biphoton state. The electric field operators at the detectors can be written in the Heisenberg picture in terms of the annihilation operators $a_{s(i)}(x_o, z_o)$ at the exit face of the SPDC crystal using the transformation [10-13]

$$E_{s(i)}^+(x_2, z_2) = \int dx_o h_{s(i)}(x_2, z_2, x_o, z_o) a_{s(i)}(x_o, z_o) \qquad (5)$$

where $h_{s(i)}$ is the impulse response for propagation over the various segments of the signal(idler) path [9,10,13]. For a free propagation path $h_{s(i)}$ is given by [14]

$$h_{s(i)}(x_2, z_2, x_o, z_o) = e^{i\pi(x_2-x_o)^2/\lambda(z_2-z_o)}. \qquad (6)$$

The correlated counting rate in equation 4 is then given by

$$CCR(x_{mi}, x_{ms}) = <\psi(x_o, z_o)| \int dx'_{oi} h^*(x_{mi}, z_{mi}, x'_{oi}, z_o) a^\dagger_i(x'_{oi}, z_o)$$

$$\times \int dx'_{os} \, h^*(x_{ms}, z_{ms}, x'_{os}, z_o) a^\dagger{}_s(x'_{os}, z_o) \int dx_{os} \, h(x_{ms}, z_{ms}, x_{os}, z_o) a_s(x_{os}, z_o)$$

$$\times \int dx_{oi} \, h(x_{mi}, z_{mi}, x_{oi}, z_o) a_s(x_{oi}, z_o) | \psi(x_o, z_o) > \quad (7)$$

where the integrals are taken over the exit face of the crystal, $z_o$ is the plane of the exit face of the crystal and $|\psi(x_o, z_o)>$ is the biphoton state at the exit face of the crystal.

The signal and idler creation operators at the exit face of the SPDC crystal can be written as

$$a^\dagger_{s(i)}(x_o, z_o) = \int d\kappa_s \, e^{i\kappa_{s(i)} x_o} e^{i\varphi_{s(i)}} a^\dagger_{s(i)}(\kappa_{s(i)}, \omega_{s(i)}, \varphi_{s(i)}) \quad (8)$$

where $a^\dagger_{s(i)}(\kappa_s, \omega_s, \varphi_s)$ is the creation operator for a signal(idler) photon in a state with transverse k-vector $\kappa_{s(i)}$, phase $\varphi_{s(i)}$ and frequency $\omega_{s(i)}$,

For the analysis in this paper we will assume that the pump beam has a Gaussian profile, $I(x) = I_o \exp(-x^2/a_P^2)$, with its waist at the center of the crystal and a diameter that is sufficiently large that its range of transverse k-vectors, $\kappa_P$, is small compared to the allowed range of $\kappa_s$ and $\kappa_i$ as set by phase matching. For the results presented in the body of the paper we will assume that the SPDC interaction is effectively phase matched. We will examine the effects of phase mismatch on the point spread functions in Appendix A.

We can then write

$$k_p = k_{pz}. \quad (9)$$

and

$$\kappa_s + \kappa_i = 0 \quad (10)$$

We will further assume narrow band degenerate operation of the SPDC so that

$$\omega_s + \omega_i = \omega_P \quad (11)$$

Finally the phases of the signal and idler k-states are related by

$$\varphi_s + \varphi_i = \varphi_P \quad (12)$$

where $\varphi_p$ is the phase of the pump which is assumed to be constant. Equations 10, 11 and 12 produce entanglement of the signal and idler through the delta functions $\delta(\kappa_s + \kappa_i)$, $\delta(\omega_s + \omega_i - \omega_P)$, and $\delta(\varphi_s + \varphi_i - \varphi_P)$.

With these assumptions and relations, the biphoton state at the exit face of the SPDC crystal can be written as [9-12]

$$|\psi(x_o, z_o)\rangle = |0\rangle + \eta \int dx_o A_P(x_o) a_s^\dagger(x_o, z_o) a_i^\dagger(x_o, z_o)|0\rangle \quad (13)$$

where $|0\rangle$ is the vacuum state and $A_P(x_o)$ is the pump amplitude distribution.

## III. RESULTS

We will calculate the correlated counting rate for the signal and idler photons in the non-collapse model as well as the detection rate for the signal photon in two collapse models. Counting rate distributions are calculated for combinations of the idler detector in the slit plane and the focal plane of the collector lens and for the signal detector in the ghost image plane and the diffraction plane. The parameters used for the calculations (as defined in Fig. 1) are $M = 2$, $I_P(x) = I_0 \exp(-x^2/a_P^2)$, $a_P = 1.5$ mm, $\lambda = 0.7022$ μm, $d_2 = 149$ mm, $f = 1000$ mm, $d_3 = 2149$ mm, $w = 160$ μm, $f_C = 50$ mm, $d_1 = 2851$ mm and $S_o = 1500$ mm.

### A. Non-Collapse Evolution

The operator product for the idler detector at the slit plane $z_{1i}$ and the signal detector in a measurement plane at $z_{ms}$ in the non-collapse model is

$$E_s^+(x_{ms}, z_{ms}) E_i^+(x_{1i}, z_{1i}) = e^{i\pi x_{1i}^2 \left(1 - M\frac{d_1}{d_2}\right)/\lambda_i S_0} \int dx_{oi} e^{-\frac{i\pi x_{oi}^2}{\lambda_i d_2}} e^{-\frac{i2\pi M x_{oi} x_{1i}}{\lambda_i d_2}}$$

$$\times e^{i\pi x_{ms}^2 /\lambda_s z_{ms}} \int dx_{os} e^{\frac{i\pi x_{os}^2}{\lambda_s z_{ms}}} e^{-\frac{i2\pi x_{os} x_{ms}}{\lambda_s z_{ms}}} a_s(x_{os}, z_o) a_i(x_{oi}, z_o) \quad (14)$$

The corresponding operator product for the idler detector in the focus of the collector lens ($x_{D1}$, $z_{fc}$) and the signal detector in a measurement plane $z_{ms}$ is

$$E_s^+(x_{ms}, z_{ms}) E_i^+(x_{D1}, z_{fc}) = \int_{-w/2}^{w/2} dx_{1i} \, e^{-i2\pi x_{1i} x_{D1}/\lambda f_c} e^{i\pi x_{1i}^2 \left(1 - M\frac{d_1}{d_2}\right)/\lambda_i S_0} \int dx_{oi} e^{-\frac{i\pi x_{oi}^2}{\lambda_i d_2}} e^{-\frac{i2\pi M x_{oi} x_{1i}}{\lambda_i d_2}}$$

$$\times e^{i\pi x_{ms}^2 /\lambda_s z_{ms}} \int dx_{os} e^{\frac{i\pi x_{os}^2}{\lambda_s z_{ms}}} e^{-\frac{i2\pi x_{os} x_{ms}}{\lambda_s z_{ms}}} a_s(x_{os}, z_o) a_i(x_{oi}, z_o) \quad (15)$$

In deriving equations 14 and 15 we used successive application of the impulse response function $h(x_j, z_j, x_k, z_k)$ for the various legs in the signal and idler paths as described in [14], and eliminated the resulting integral over the lens planes using the method of stationary phase (see Appendices B and C).

The correlated counting rates for various combinations of the idler detector in the slit plane or the focus of the collector lens and the signal at the ghost image plane or the diffraction plane are calculated using equations 7, 13, 14 and 15 and the commutator

$$\left[a_{s(i)}(x'_{os(i)}, z_o), a_{s(i)}^\dagger(x_o, z_o)\right] = \delta(x'_{os(i)} - x_o). \quad (16)$$

The expressions for the correlated counting rates for the various combinations of detector positions are

Integrating idler detector in slit plane, point signal detector in ghost image plane with coordinates $x_{ms} = x_{2s}$ and $z_{ms} = d_2$

$$CCR(z_{1i}, x_{2s}, d_2) = \int_{-w/2}^{w/2} dx_{1i} |f(x_{1i}, z_{1i}, x_{2s}, d_2)|^2 \tag{17}$$

Integrating idler detector in slit plane, point signal detector in diffraction plane with coordinates $x_{ms} = x_{3s}$ and $z_{ms} = d_3$

$$CCR(z_{1i}, x_{3s}, d_3) = \int_{-w/2}^{w/2} dx_{1i} |f(x_{1i}, z_{1i}, x_{3s}, d_3)|^2 \tag{18}$$

Integrating idler detector in collector lens focal plane, point signal detector in ghost image plane

$$CCR(f_c, x_{2s}, d_2) = \int dx_{D1} |g(x_{D1}, f_c, x_{2s}, d_2)|^2 \tag{19}$$

Integrating idler detector in collector lens focal plane, point signal detector in diffraction plane

$$CCR(f_c, x_{3s}, d_3) = \int dx_{D1} |g(x_{D1}, f_c, x_{3s}, d_3)|^2 \tag{20}$$

where

$$f(x_{1i}, z_{1i}, x_{2s}, d_2) = e^{i\pi x_{2s}^2/\lambda d_2} \int dx_{os} e^{-\frac{x_{os}^2}{2a_P^2}} e^{-\frac{i2\pi x_{os}[Mx_{1i}+x_{2s}]}{\lambda d_2}} \tag{21}$$

$$f(x_{1i}, z_{1i}, x_{3s}, d_3) = e^{i\pi x_{3s}^2/\lambda d_3} \int dx_{os} e^{-\frac{x_{os}^2}{2a_P^2}} e^{\frac{i\pi x_{os}^2(d_2-d_3)}{\lambda d_2 d_3}} e^{-\frac{i2\pi x_{os}\left[Mx_{1i}+x_{3s}\left(\frac{d_2}{z_3}\right)\right]}{\lambda d_2}} \tag{22}$$

$$g(x_{D1}, f_c, x_{2s}, d_2) = \int_{-w/2}^{w/2} dx_{1i} e^{-\frac{i2\pi x_{1i}x_{D1}}{\lambda f_c}} f(x_{1i}, z_{1i}, x_{2s}, d_2) \tag{23}$$

$$g(x_{D1}, f_c, x_{3s}, d_3) = \int_{-w/2}^{w/2} dx_{1i} e^{-\frac{i2\pi x_{1i}x_{D1}}{\lambda f_c}} f(x_{1i}, z_{1i}, x_{3s}, d_3) \tag{24}$$

### 1. Results for Idler Detector in Slit Plane

We consider first the evolution of the system for the idler detector in the slit plane and the signal detector in the ghost image and diffraction planes in the Heisenberg picture. For this analysis the relevant correlated counting rates are CCR($z_{1i}$, $x_{2s}$, $d_2$) and CCR($z_{1i}$, $x_{3s}$, $d_3$) in equations 17 and 18. These are shown in Figs. 2 and 3.

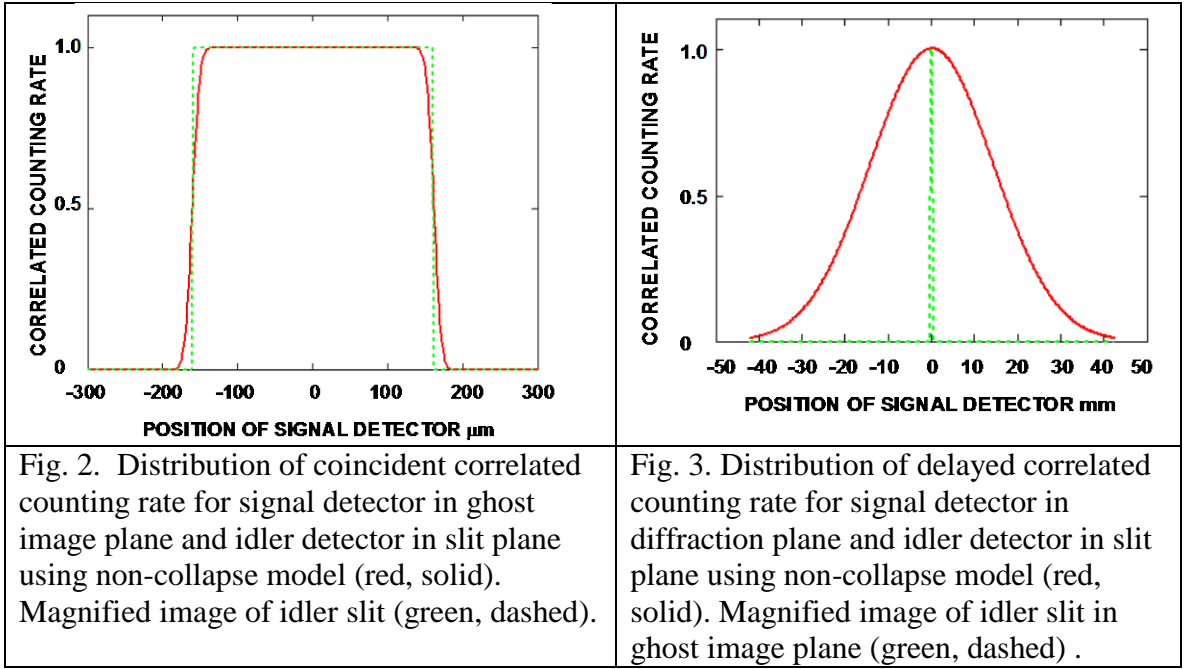

| Fig. 2. Distribution of coincident correlated counting rate for signal detector in ghost image plane and idler detector in slit plane using non-collapse model (red, solid). Magnified image of idler slit (green, dashed). | Fig. 3. Distribution of delayed correlated counting rate for signal detector in diffraction plane and idler detector in slit plane using non-collapse model (red, solid). Magnified image of idler slit in ghost image plane (green, dashed). |
|---|---|

The correlated counting rate distribution for the signal in the ghost image plane shows a reasonably well resolved magnified image of the idler slit. The distribution in the diffraction plane is wider as a result of spreading after the ghost image plane.

### B. Collapse Evolution

We now consider the concept of wavefunction collapse effected by the idler measurement. In this concept the signal and idler evolve under the Heisenberg picture until the idler reaches the slit plane and the signal reaches the ghost image plane. At this time the idler is measured, projecting the biphoton state into a single un-entangled signal quantum state whose properties depend on the result of the idler measurement. The projected signal state then serves as an initial condition for further evolution within the system. In our analysis, the collapsed signal state will propagate from the ghost image plane to the diffraction plane where it will be detected without further correlation since the idler has been destroyed in the (earlier) measurement and there is no longer an entangled biphoton state.

### 1. Pure-State Collapse

In order to calculate the distribution of the signal in the diffraction plane it is necessary to choose a model for the description of the collapsed state. We consider first the state described in Ref. 8. There it is stated that the idler measurement localizes the idler state at the time of measurement to a transverse dimension the size of the slit, establishing an idler wave function of the form [15]

$$\psi_i(x_{1i}) = \begin{matrix} 1 & |x_{1i}| \leq w/2 \\ 0 & |x_{1i}| > w/2 \end{matrix} \qquad (25)$$

The collapsed signal state is formed by integrating the inner product of the biphoton state at the idler slit plane and the ghost image plane and the idler wave function established by the measurement of the idler over the idler slit, since the measurement integrates the idler that is transmitted through the slit. The resulting collapsed signal state at the ghost image plane is given by

$$|\psi_s(x_{2s})> = \int_{-\frac{w}{2}}^{\frac{w}{2}} dx_{1i} f(x_{1i}, z_{1i}, x_{2s}, d_2) \; |1s> \qquad (26)$$

where $f(x_{1i}, z_{1i}, x_{2s}, d_2)$ is defined in equation 21. The counting rate distribution of the collapsed signal state in the ghost image plane is given by

$$CR_{coll}(x_{2s}, d_2) = <\psi_s(x_{2s}, d_2)|a_s^\dagger a_s|\psi_s(x_{2s}, d_2)>$$

$$= |\int_{-\frac{w}{2}}^{\frac{w}{2}} dx_{1i} f(x_{1i}, z_{1i}, x_{2s}, d_2)|^2 \qquad (27)$$

Upon propagation to the diffraction plane, the collapsed signal state takes on the form

$$|\psi_s(x_{3s})> = \int dx_{2s} \, e^{i\pi(x_{3s}-x_{2s})^2/\lambda(d_3-d_2)} \int_{-\frac{w}{2}}^{\frac{w}{2}} dx_{1i} f(x_{1i}, z_{1i}, x_{2s}, d_2) \; |1s> \qquad (28)$$

The counting rate distribution for the collapsed signal state in the diffraction plane is given by

$$CR_{coll}(x_{3s}, d_3) = <\psi_s(x_{3s}, d_3)|a^\dagger a|\psi_s(x_{3s}, d_3)>$$

$$= |\int_{-\frac{w}{2}}^{\frac{w}{2}} dx_{1i} f(x_{1i}, z_{1i}, x_{3s}, d_3)|^2 \qquad (29)$$

The counting rate distributions for the collapsed signal state in the ghost image plane and the diffraction plane are shown in Figs. 4 and 5.

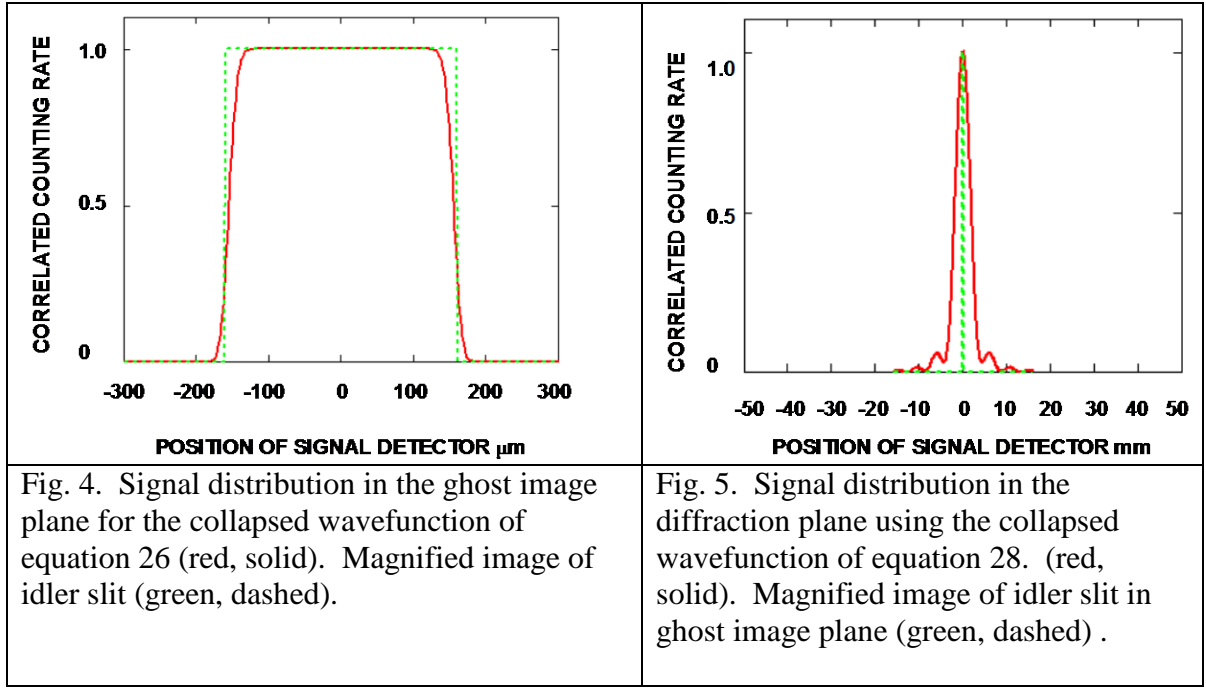

Fig. 4. Signal distribution in the ghost image plane for the collapsed wavefunction of equation 26 (red, solid). Magnified image of idler slit (green, dashed).

Fig. 5. Signal distribution in the diffraction plane using the collapsed wavefunction of equation 28. (red, solid). Magnified image of idler slit in ghost image plane (green, dashed) .

The signal distribution in the ghost image plane using the collapsed quantum state of equation 26 gives an approximate image of the idler slit and is very similar to the distribution in Fig. 2 using the non-collapse Heisenberg calculation . However, comparison of Figs. 5 and 3 shows that the distribution in the diffraction plane is much narrower for the collapsed wavefunction of equation 28 than that of the non-collapse Heisenberg model.

## 2. Mixed State Collapse

We now consider an alternative collapse model involving a mixed state for the signal. This model is motivated by a consideration of the details of the photoelectric detection of the idler and finds support in Refs. 5 and 6. To examine the role of the photoelectric detection of the idler in giving rise to the signal collapse, we model the integrating idler detector as an array of point detectors. The outputs of the individual sensors in the array can be connected in parallel to model a single large area detector, or can be recorded individually in memory and integration can be done in post processing. When an idler photon is detected at the slit, one and only one of the detectors in the array will respond. It is this response that causes the signal state to collapse at the ghost image plane, and that collapsed state will then propagate to the diffraction plane where it will be detected by detector D2. Since we are considering a low flux regime we can expect that the collapse, propagation and detection of one signal photon will all occur before a second signal idler pair is generated. Thus we would expect that the counting rate in the diffraction plane at a given position will be the incoherent sum of the counting rates for the individual collapsed states at that position. A given collapsed state, in turn, will be exactly the state left from the biphoton state whose idler component was detected. The mode function of this state is the point spread function at the ghost plane associated with detection of an idler photon at position $x_{1i}$.

We thus consider that the collapsed signal state is a mixed state described by the density operator

$$\hat{\rho} = |\psi_n > P_n < \psi_n. \tag{30}$$

The basis states of the mixed state are the point spread functions associated with an idler detection at position $x_{1i}$ given by

$$|\psi_n(x_{2s}, d_2) > = |\psi_{1i}(x_{2s}, d_2) > = K_{1i} f(x_{1i}, z_{1i}, x_{2s}, d_2)|1s> \tag{31}$$

The normalizing constant $K_{1i}$ is given by

$$K_{1i}^2 \int dx_{2s} |f(x_{1i}, z_{1i}, x_{2s}, d_2)|^2 = 1 \tag{32}$$

The $P_n$ are the diagonal elements of the density matrix which has the form

$$\rho = \begin{pmatrix} P_1 & \cdots & 0 \\ \vdots & \ddots & \vdots \\ 0 & \cdots & P_n \end{pmatrix}. \tag{33}$$

The density matrix element $P_n$ is the probability of an idler detection at position $x_n = x_{1i}$ given by

$$P_n = P_{1i} = \int dx_{2s}| f(x_{1i}, z_{1i}, x_{2s}, d_2)|^2, \tag{34}$$

and the total probability is normalized to unity

$$\int dx_{1i} P_{1i} = K_2 \int dx_{1i} \int dx_{2s}| f(x_{1i}, z_{1i}, x_{2s}, d_2)|^2 = 1 \tag{35}$$

The counting rate distribution for the collapsed signal in the ghost image plane is given by

$$CR_{collmixed}(x_{2s}, d_2) = tr(\hat{\rho} \, a_s^\dagger a_s) \tag{36}$$

where the trace is taken over the point spread functions of equation 31. The counting rate in the diffraction plane is obtained using equation 36 with the trace taken over the point spread functions after they have propagated to the diffraction plane, i. e. over the set of functions

$$|\psi_n(x_{3s}, d_3) > = |\psi_{1i}(x_{3s}, d_3) > = K_{1i} f(x_{1i}, z_{1i}, x_{3s}, d_3)|1s> \tag{37}$$

The counting rate distributions in the ghost-image and diffraction planes for a single basis state at the position $x_{1i} = 0$ in the idler slit plane is shown in Fig. 6 in the ghost image plane.

| 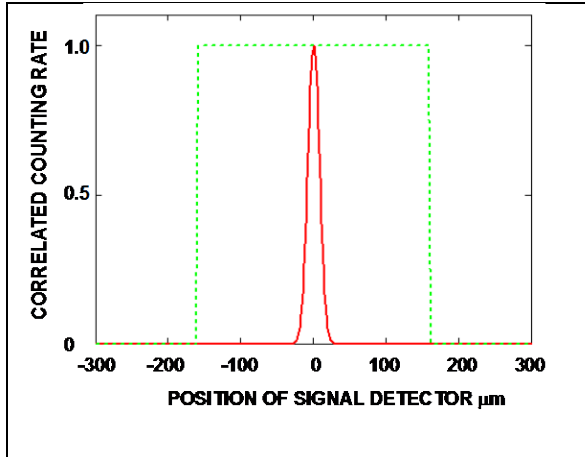 | 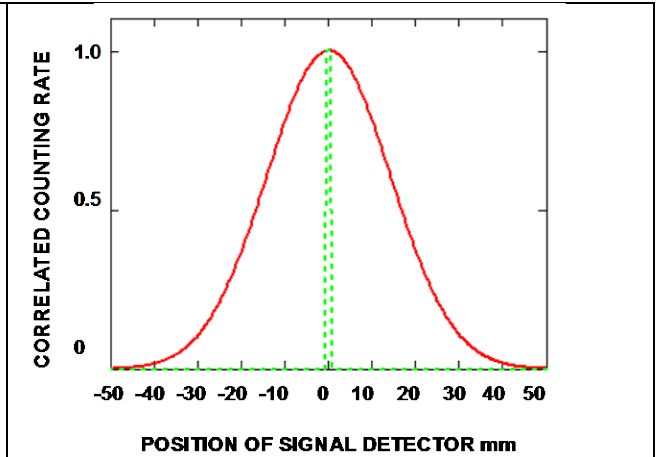 |
|---|---|
| Fig. 6. The counting rate distribution in the ghost image plane for a single basis state at $x_{1i} = 0$ in the mixed state collapse model (red, solid). Magnified image of idler slit (green, dashed). | Fig. 7. The counting rate distribution for a single collapsed state at $x_{1i} = 0$ in the diffraction plane (red, solid). Magnified image of idler slit in ghost image plane (green, dashed). |

It can be seen that the distribution in the ghost image plane for a single basis state (Fig. 6) is much narrower than the magnified image of the slit. This is to be expected since the basis state is derived from the point spread function of the biphoton and we have chosen a sufficiently large numerical aperture, which is determined by $a_P/d_2$, so that the imaging resolution, which is determined by the point spread function, is substantially smaller than the width of the slit.

The width of the distribution of the counting rate in the diffraction plane from the single collapsed state in the mixed state model is seen to be comparable to the width of the correlated counting rate in the Heisenberg model (Fig. 3), and considerably wider that the width of the distribution from the pure state collapse model (Fig. 5).

We next consider the distributions in the ghost imaging plane and the diffraction plane for basis functions at different positions $x_{1i}$ in the idler slit (Figs. 8 and 9)

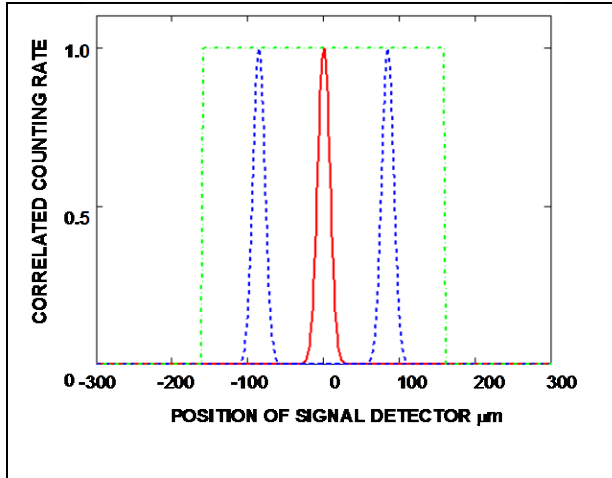 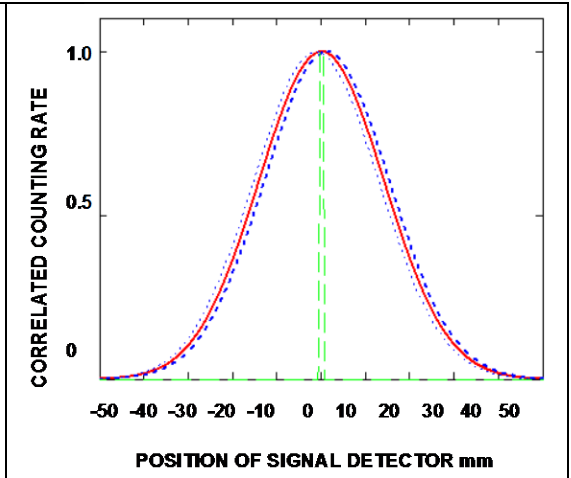

Fig. 8. Counting rate distributions in the ghost image plane for basis functions at $x_{1i} = 0$ (red, solid) and $x_{1i} = \pm 42$ μm (blue, dashed). Magnified image of idler slit (green, dashed).

Fig. 9. Counting rate distributions in the diffraction plane for basis functions at $x_{1i} = 0$ (red, solid) and $x_{1i} = \pm 42$ μm (blue, dashed, dotted). Magnified image of idler slit in ghost image plane (green, dashed)..

The counting rate distributions for basis functions corresponding to idler detections at different values of $x_{1i}$ are seen to be narrower than the width of the idler slit, but to be centered on different transverse positions, as would be expected since their location is related to an image of the position of the idler detection within the idler slit. The widths of the distributions in the diffraction plane from the individual basis states are also seen to be similar to each other. They are each of approximately the same width as that of the Heisenberg evolution in the diffraction plane (Fig. 3) and only minimally separated in space.

      The total counting rate for the signal in the mixed state collapse model is obtained by integrating over the counting rate distributions of the individual basis function with each basis function weighted by its probability of detection at the idler slit. The distribution of detection probability, shown in Fig. 10, is constant over the slit.

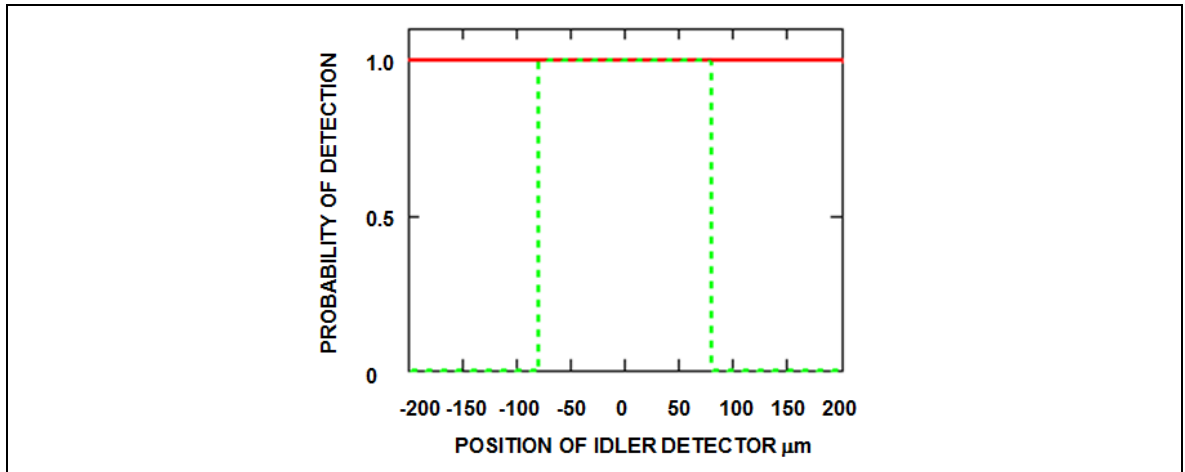

Fig. 10 Probability distribution of idler detection in the slit plane

The resulting counting rate distributions are shown in Figs. 11 and 12 for the ghost image plane and diffraction plane.

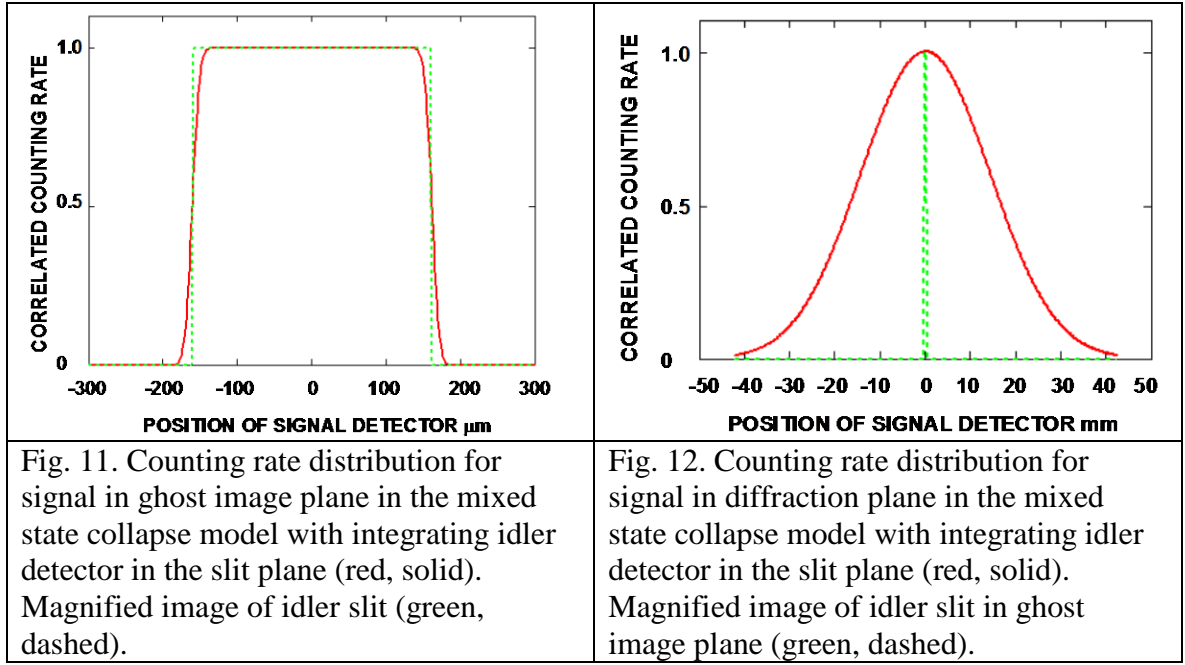

| Fig. 11. Counting rate distribution for signal in ghost image plane in the mixed state collapse model with integrating idler detector in the slit plane (red, solid). Magnified image of idler slit (green, dashed). | Fig. 12. Counting rate distribution for signal in diffraction plane in the mixed state collapse model with integrating idler detector in the slit plane (red, solid). Magnified image of idler slit in ghost image plane (green, dashed). |

Comparison of Figs. 11 and 12 with Figs. 2 and 3 show that the distributions for the mixed state collapse model and the Heisenberg model are identical in both the ghost image plane and the diffraction plane. While the counting rate distributions of the individual basis functions of the mixed state collapse in the ghost image plane are each narrow, they are centered at different positions within the region of the image of the idler slit depending on the location of the idler photon detection. It is the incoherent sum of the individual counting rate distributions in the mixed state collapse model that reproduces the image of the idler slit. In the diffraction plane, the counting rate distributions for each of the individual basis functions are nearly identical and the counting rate distribution in the diffraction plane for the incoherent sum is only minimally different from that of the individual basis functions.

### C. Idler Detector In Lens Collection Plane

The equivalent calculations can be done when the idler detector is placed at the focus of the collecting lens. The results are shown in Figs 13 -20

| | |
|---|---|
| 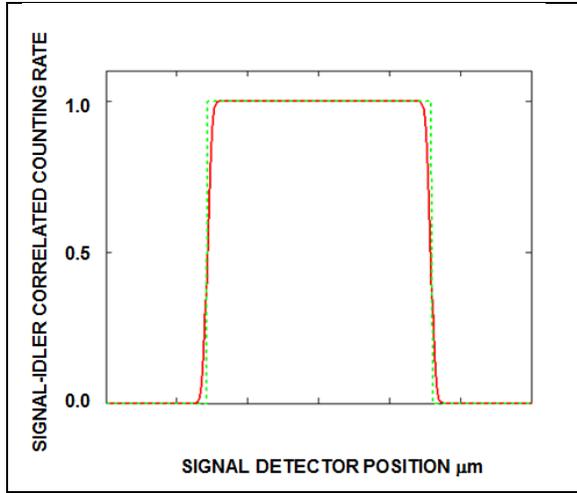 Fig. 13 Distribution of coincident correlated counting rate for signal detector in ghost image plane and idler detector in the focal plane of the collector lens using the non-collapse model (red, solid). Magnified image of idler slit (green, dashed). | 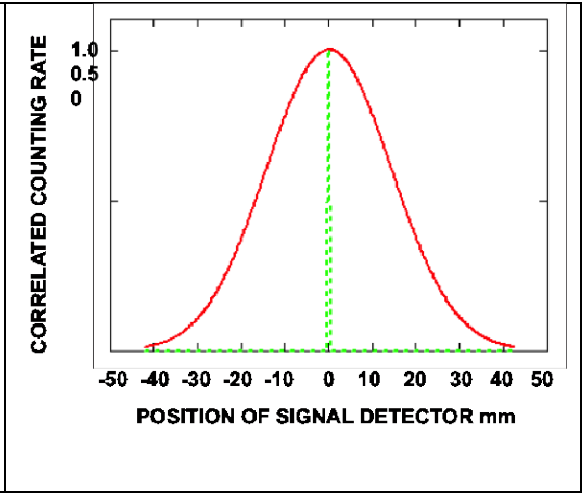 Fig. 14. Distribution of delayed correlated counting rate for signal detector in diffraction plane and idler detector in the focal plane of the collector lens using the non-collapse model (red, solid). Magnified image of idler slit in ghost image plane (green, dashed). |
| 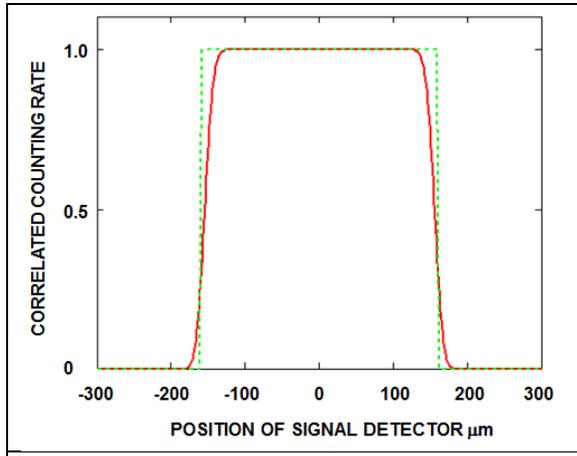 Fig. 15. The signal counting rate distribution in the ghost image plane for a single basis state at $xD1 = 0$ in the mixed state collapse model (red, solid). Magnified image of idler slit (green, dashed). | 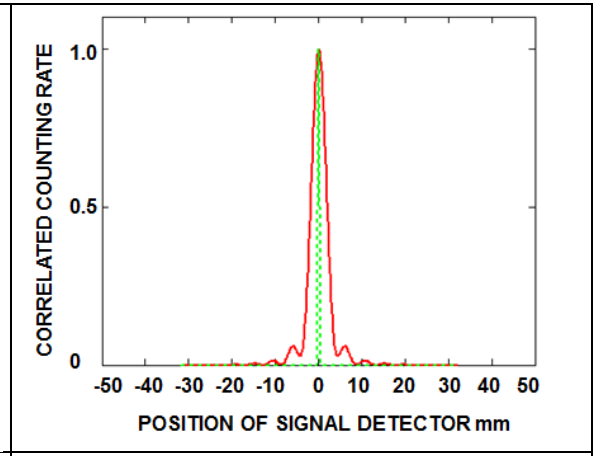 Fig. 16. The signal counting rate in the diffraction plane for a single basis state at $xD1 = 0$ in the mixed state collapse model (red, solid). Magnified image of idler slit in ghost image plane (green, dashed). |

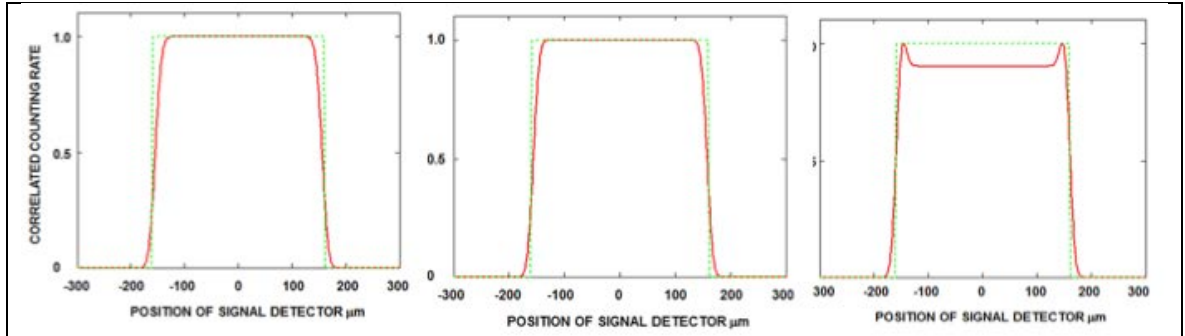
Fig. 17. The signal counting rate distributions in the ghost image plane for basis states in the mixed state collapse model at xD1 = 0 (left), 0.25 (center) and 0.5 (right) (red, solid). Magnified image of idler slit (green, dashed).

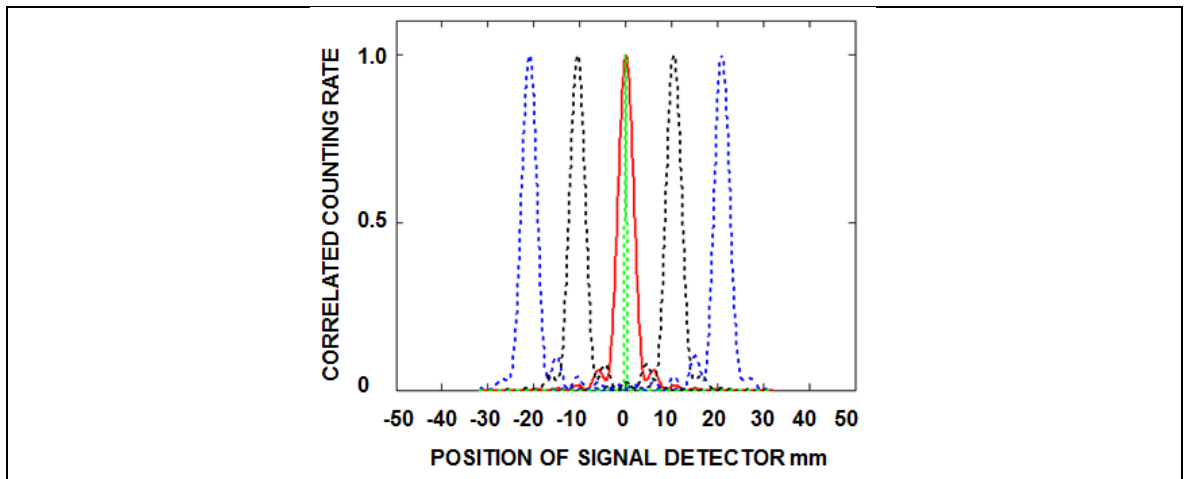
Fig. 18 The signal counting rate distributions in the diffraction plane for basis states in the mixed state collapse model at xD1 = 0 (red, solid), ±0.25 (black, dashed) and ±0.5 (blue, dashed). Magnified image of idler slit in ghost image plane (green, dashed).

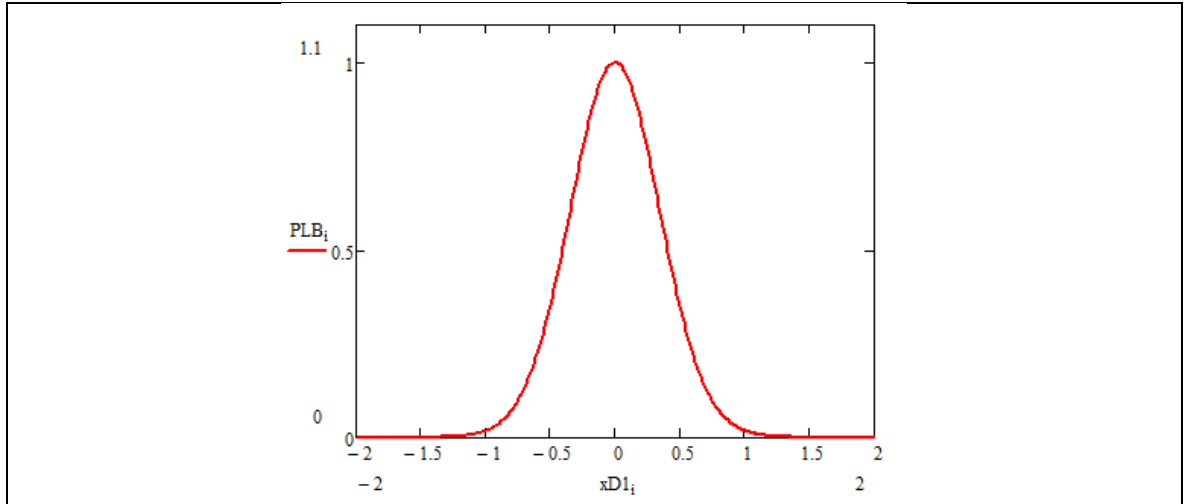

Fig. 19. Probability of idler detection (density matrix elements) for idler detector in focal plane of collector lens in mixed state collapse model.

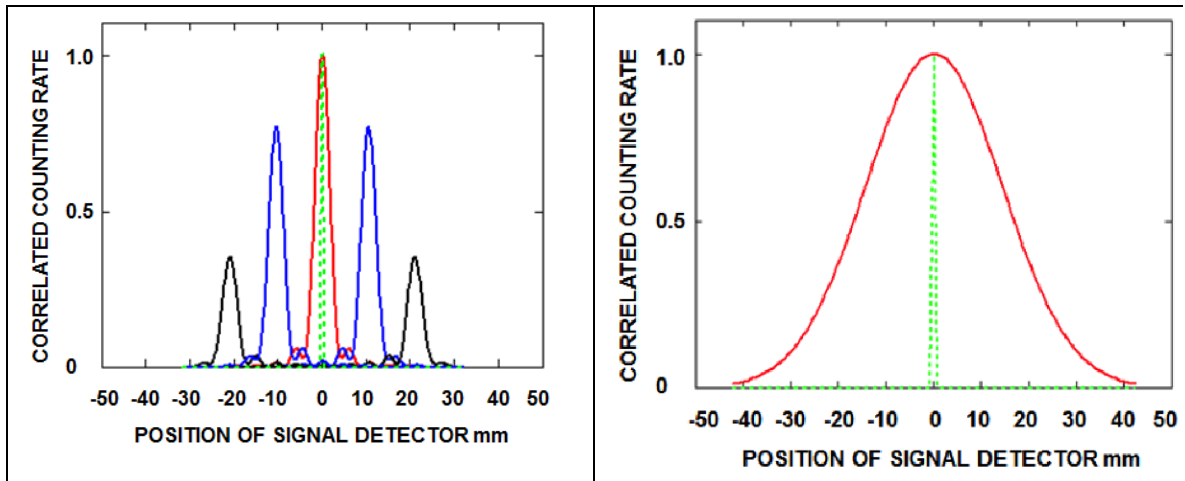

| Fig. 20. Basis states of Fig. 18 in the diffraction plane weighted by density matrix elements for the idler detector in the focal plane of the collector lens. Magnified image of idler slit in ghost image plane (green, dashed). | Fig. 21. Integrated signal counting rate distribution in diffraction plane for mixed state collapse. Magnified image of idler slit in ghost image plane (green, dashed). |

The overall conclusion for the correlated counting rates when the idler detector is located in the focal plane of the collecting lens is that the non-collapse Heisenberg predictions and the predictions of the mixed state collapse model are in agreement, as can be seen by comparing Figs. 3 and 21. However, the details of the evolution are somewhat different from when the idler detector is in the slit plane. When the idler detector is at the focus of the collecting lens, each of the individual basis functions forms an image of the slit in the ghost plane (Fig. 17). The distributions in the diffraction plane from the individual basis functions are considerably narrower than the distribution predicted by the Heisenberg evolution (Fig. 18), just as in the pure state collapse of Fig. 5. However, in the mixed state collapse model, the diffraction patterns from different basis functions are centered at different positions in the diffraction plane, depending on the location of the idler detection in the collecting lens focal plane. When the total counting rate distribution is calculated as the sum of the counting rates from the individual basis functions, weighted by the density matrix distribution in Fig. 19, it is found to be equal to the counting rate distribution of the Heisenberg evolution.

## IV. ANALYTIC COMPARISON OF THE NON-COLLAPSE (HEISENBERG) MODEL AND THE MIXED STATE COLLAPSE MODEL

The predictions of the mixed state collapse model and the Heisenberg model for the counting rate distribution in the diffraction plane can be compared analytically.

The counting rate in the diffraction plane in the mixed state collapse model is given by

$$CR_{collmixed}(x_{3s}, d_3) = \int d\,xD1 P_{xD1} |\psi_{xD1}(x_{3s}, d_3)|^2 = K_1^2 \int d\,xD1 P_{xD1} |g(xD1, x_{3s}, d_3)|^2 \quad (35)$$

where

$$P_{xD1} = K_2 \int dx_{2s}\,|g(xD1, x_{2s}, d_2)|^2 \quad (36)$$

$$K_1^2 \int dx_{2s}|\,g(xD1, x_{2s}, d_2)|^2 = 1 \quad (37)$$

$$P_{xD1} = K_2 \int dxD1\,|g(xD1, x_{2s}, d_2)|^2 \quad (38)$$

$$K_2 \int dxD1 \int dx_{2s}\,|g(xD1, x_{2s}, d_2)|^2 = 1 \quad (39)$$

$$CR_{collmixed}(x_{3s}, d_3) = K_2 \int d\,xD1 K_1^2 \int dx_{2s}\,|g(xD1, x_{2s}, d_2)|^2 |g(xD1, x_{3s}, d_3)|^2 \quad (40)$$

$$CR_{collmixed}(x_{3s}, d_3) = K_2 \int d\,xD1 |g(xD1, x_{3s}, d_3)|^2 = K_2 CR_{Heisenberg}(x_{3s}, d_3) \quad (41)$$

where the last equality appears as a result of equation 17. Thus, apart from a normalizing factor, the analytic form of the counting rate distribution in the diffraction plane using the mixed state collapse model is identical to that of the non-collapse Heisenberg model.

## V. SUMMARY

The evolution of an entangled signal idler pair in the geometry of ghost imaging is described with collapse and non-collapse models. Two collapse models are considered. One collapse model reduces the signal to a pure state as described in Ref 7. The second collapse model reduces the signal to a mixed state. The mixed state collapse model gives predictions of the counting rates in different propagation planes and idler detection planes that are in agreement with the corresponding predictions of the non-collapse model. However, the predictions of both the non-collapse and the mixed state collapse model are substantially different from the prediction of the pure state collapse model when diffraction away from the collapse plane is considered. The predicted differences offer the possibility of experimental distinction between the two collapse models. The calculations of the mixed state model also indicate that differences that can arise in state preparation when substantial propagation paths are involved and information is encoded in the wavefront in addition to the polarization. In such a situation, it makes a difference as to whether the prepared state can be distinguished in the near and far field depending on just how the entangled idler is measured.

## ACKNOWLEDGEMENTS

This work was supported by the Office of Naval Research.

# APPENDIX A: EFFECT OF PHASE MISMATCH

The analysis in the main body of the paper assumed perfect phase matching and demonstrated that the image resolution in entangled ghost imaging is limited by the pump beam aperture. In a general case, the results of the main text can be affected by a phase mismatch caused by the combination of the size of the pump beam aperture and the length of the SPDC crystal. In this appendix we examine the effect of phase mismatch on the point spread function in the ghost image plane that form the basis states for the mixed state collapse model.

The geometry of the phase mismatch with and without transverse pump components is illustrated in Fig. A1.

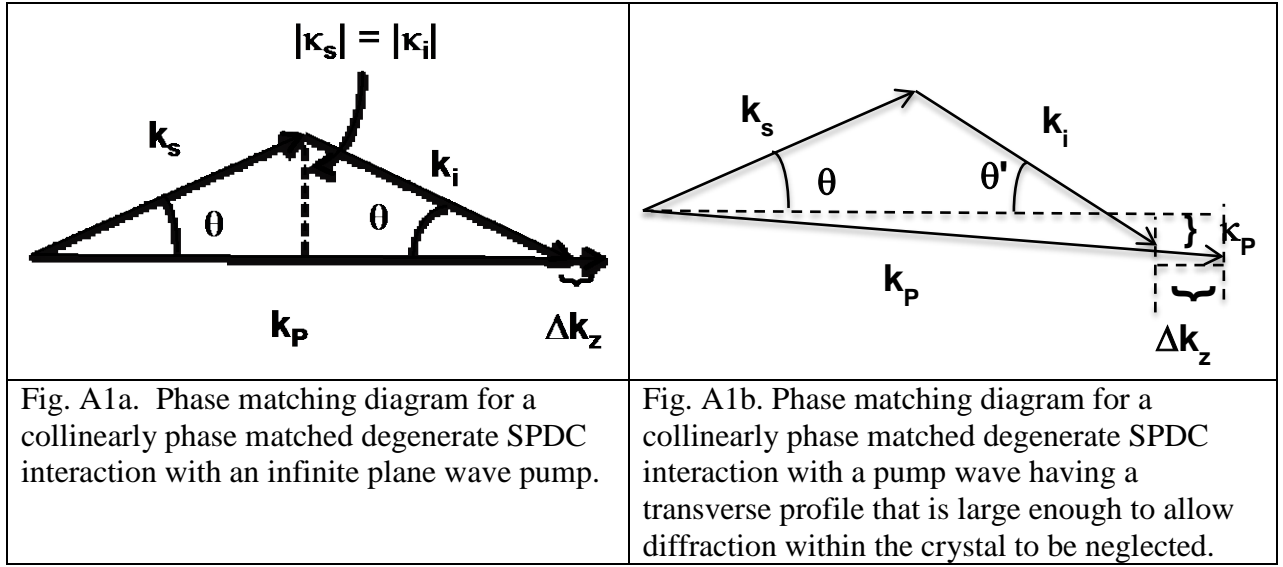

| Fig. A1a. Phase matching diagram for a collinearly phase matched degenerate SPDC interaction with an infinite plane wave pump. | Fig. A1b. Phase matching diagram for a collinearly phase matched degenerate SPDC interaction with a pump wave having a transverse profile that is large enough to allow diffraction within the crystal to be neglected. |
|---|---|

Phase mismatch in the z (forward) direction for an infinite plane wave pump in an SPDC interaction that is collinearly phase matched is illustrated in Fig. A1a. For the degenerate operation considered in this paper, the phase mismatch arises for off-axis signal and idler beams and ultimately limits their angular content. In this situation the transverse components of the signal and idler k vectors are equal and opposite, i. e. $\kappa_s = -\kappa_i$. The phase mismatch geometry for a pump beam with transverse k-vector component $\kappa_P$ is illustrated in Fig. A1b. In this situation the transverse components of the k-vectors satisfy the relation $\kappa_s = \kappa_P - \kappa_i$.

The correlated counting rate for a point idler detector in the slit plane and a signal detector in the ghost image plane is given by

$$CCR(0, x_{2s}, d_2) = \left| <0 \right| \int dx_{os} \int dx_{oi} e^{\frac{i\pi(x_{os}^2 - x_{oi}^2)}{\lambda_i d_2}} e^{-\frac{i2\pi x_{os} x_{2s}}{\lambda_i d_2}} \int dx \int d\kappa_P \int d\kappa_s \int d\kappa_i$$

$$\times \int d\kappa_s' \int d\kappa_i' A_P(\kappa_P) \operatorname{sinc}\left(\frac{\Delta k_z L}{2}\right) e^{i\Delta k_z L/2} e^{i\kappa_P x} e^{-i\kappa_s x} e^{-i\kappa_i x} e^{i\kappa_s' x_{os}} e^{i\kappa_i' x_{oi}}$$

$$\times \, a_s(\kappa'_s) a_s(\kappa'_i) a_s^\dagger(\kappa_s) a_i^\dagger(\kappa_i) |0> \bigg|^2 \tag{A1}$$

where L is the length of the SPDC crystal, $A_P(\kappa_P)$ is the transverse Fourier distribution of the pump amplitude, $\Delta k_z$ is the wavevector mismatch in the z direction and

$$sinc(x) = \frac{sinx}{x}$$

Using the commutator

$$[a_{s(i)}(\kappa'_{s(i)}), a^\dagger_{s(i)}(\kappa_{s(i)})] = \delta(\kappa'_{s(i)} - \kappa_{s(i)}) \tag{A2}$$

and integrating on $\kappa'_s$ and $\kappa'_i$ gives

$$CCR(0, x_{2s}, d_2) = \bigg| \int dx_{os} \int dx_{oi} e^{\frac{i\pi(x_{os}^2 - x_{oi}^2)}{\lambda_i d_2}} e^{-\frac{i2\pi x_{os} x_{2s}}{\lambda_i d_2}} \int dx \int d\kappa_P \int d\kappa_s \int d\kappa_i$$

$$\times A_P(\kappa_P) sinc(\tfrac{\Delta k_z L}{2}) e^{i\Delta k_z L/2} e^{i\kappa_P x} e^{-i\kappa_s x} e^{-i\kappa_i x} e^{i\kappa_s x_{os}} e^{i\kappa_i x_{oi}} \bigg|^2 \tag{A3}$$

Integrating on x gives

$$CCR(0, x_{2s}, d_2) = \bigg| \int dx_{os} \int dx_{oi} e^{\frac{i\pi(x_{os}^2 - x_{oi}^2)}{\lambda_i d_2}} e^{-\frac{i2\pi x_{os} x_{2s}}{\lambda_i d_2}} \int d\kappa_P \int d\kappa_s \int d\kappa_i$$

$$\times A_P(\kappa_P) sinc(\tfrac{\Delta k_z L}{2}) e^{i\Delta k_z L/2} e^{i\kappa_s x_{os}} e^{i\kappa_i x_{oi}} \delta(\kappa_P - \kappa_s - \kappa_i) \bigg|^2. \tag{A4}$$

The phase mismatch is given by

$$\Delta k_z = \frac{\kappa_s^2}{2k_s} + \frac{\kappa_i^2}{2k_i} - \frac{\kappa_P^2}{2k_P} \tag{A5}$$

Integrating on $\kappa_s$ gives

$$CCR(0, x_{2s}, d_2) = \bigg| \int dx_{os} \int dx_{oi} e^{\frac{i\pi(x_{os}^2 - x_{oi}^2)}{\lambda_i d_2}} e^{-\frac{i2\pi x_{os} x_{2s}}{\lambda_i d_2}} \int d\kappa_P \int d\kappa_i$$

$$\times A_P(\kappa_P) e^{i\kappa_P x_{os}} sinc[\left(\frac{(\kappa_i - \kappa_P)^2}{4k_s} + \frac{\kappa_i^2}{4k_i} - \frac{\kappa_P^2}{4k_P}\right) L] e^{i\left(\frac{(\kappa_i - \kappa_P)^2}{4k_s} + \frac{\kappa_i^2}{4k_i} - \frac{\kappa_P^2}{4k_P}\right) L} e^{i\kappa_i(x_{oi} - x_{os})} \bigg|^2 \tag{A6}$$

Under the assumption that the pump beam diameter is large enough so that diffraction of the pump within the crystal can be neglected the range of $\kappa_P$ is much smaller than the range of $\kappa_i$ allowed by off axis phase matching. As a result we can drop the $\kappa_P$ term in the sinc function in equation A6. Setting $k_s = k_i$ and integrating on $\kappa_P$ gives

$$CCR(0, x_{2s}, d_2) =$$
$$\left| \int dx_{os} \int dx_{oi} e^{\frac{i\pi(x_{os}^2 - x_{oi}^2)}{\lambda_i d_2}} e^{-\frac{i2\pi x_{os} x_{2s}}{\lambda_i d_2}} A_P(x_{os}) \int d\kappa_i \, sinc\left(\frac{\kappa_i^2 L}{2k_i}\right) e^{i\kappa_i^2 L/2k_i} e^{i\kappa_i(x_{oi} - x_{os})} \right|^2 \quad (A7)$$

Integrating over $x_{oi}$ by stationary phase gives

$$CCR(0, x_{2s}, d_2) = \left| \int dx_{os} \, e^{\frac{i\pi x_{os}^2}{\lambda_i d_2}} e^{-\frac{i2\pi x_{os} x_{2s}}{\lambda_i d_2}} A_P(x_{os}) \int d\kappa_i \, sinc\left(\frac{\kappa_i^2 L}{2k_i}\right) e^{i\kappa_i^2 \left(\frac{\lambda_i d_2}{4\pi} + \frac{L}{2k_i}\right)} e^{-i\kappa_i x_{os}} \right|^2$$
(A8)

Setting L = 0 in the sinc function and exponential in equation A8 and integrating by stationary phase on $\kappa_i$ gives the same form as the correlated counting rate for a point idler detector at the idler slit given by $[f(0, z_{1i}, x_{2s}, d_2)]^2$ from equation 21. The sinc function and exponential term in L in equation A8 account for the effects of phase mismatch.

The effect of phase mismatch on the point spread function for the parameters of the interaction described in the main text is shown in Fig. A2.

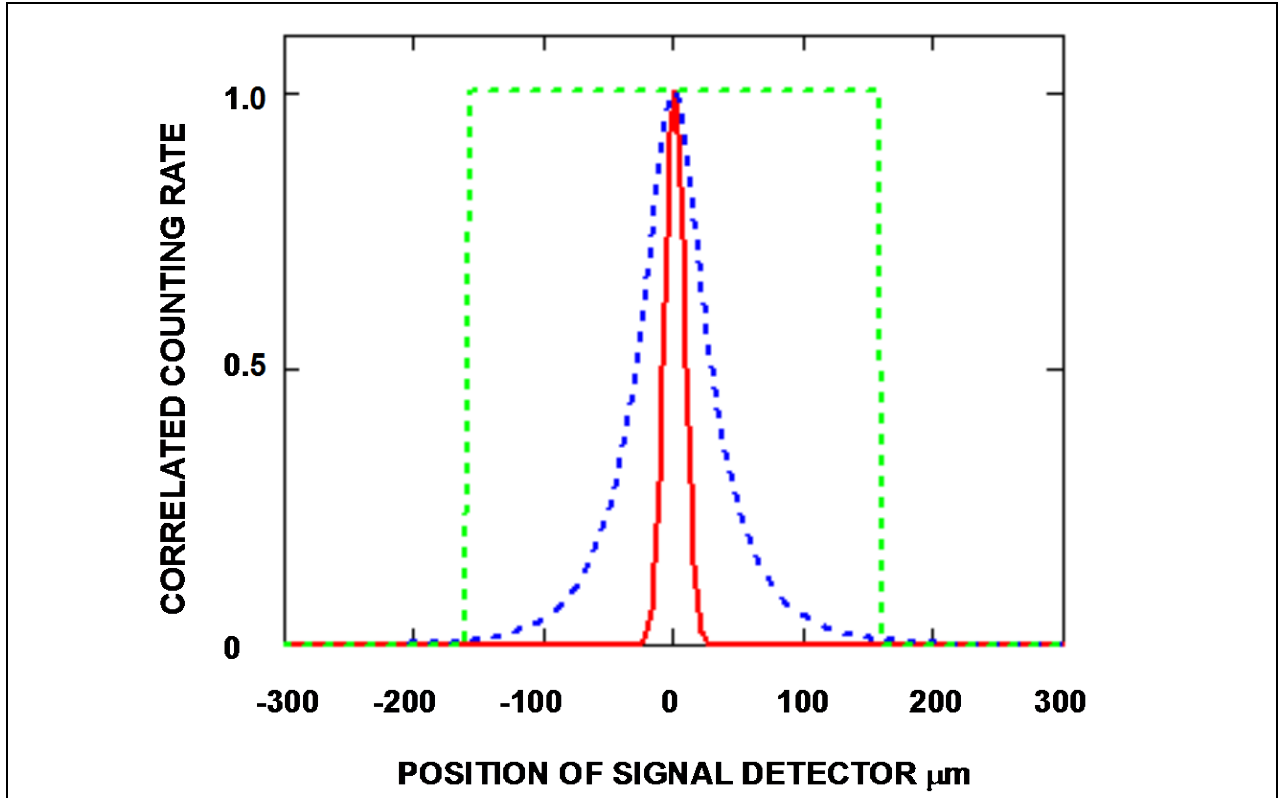

Fig. A2. Comparison of the point spread function for an idler detector in the slit plane and a signal detector in the ghost image plane with (blue dotted) and without (red solid) phase mismatch. Magnified image of idler slit (green, dashed).

It can be seen that for the parameters of the system described in the main text, the phase mismatch broadens the point spread function somewhat in the half-width and more in the tails, but does not affect the conclusions of the main paper.

Additional contributions to off-axis components can occur in nondegenerate operation as discussed in [11]. When $\omega_s \neq \omega_i$ off axis components of the signal and idler can be phase matched, and can make the dominant contribution to the correlated signal-idler count rate. In addition, the position of the ghost image plane is displaced from the one given in equation 2 by an amount $\Delta z(\theta) = [1 - (\lambda_{so}/\lambda_{s\prime})]d_2$, where $\lambda_{so} = 2\lambda_P$ and $\lambda_{s\prime}$ is the signal wavelength that is phase matched at angle $\theta$. The geometry of the ghost image configuration treated here effectively limits the off axis angles to a maximum value that is approximately given by $\theta = a_P/d_2$. For the parameters of the configuration described in this paper this angle is about 10 mrad. In BBO, the corresponding phasematched signal wavelength is $\lambda_{s\prime} = 1.002\lambda_{so}$. These effects do not change the point spread noticeably from the ones given in Fig. A2. As a result it is not expected that nondegenerate effects will alter the conclusions of the paper.

## APPENDIX B: INTEGRATION OVER THE IMAGING LENS PLANE BY STATIONARY PHASE.

Analytic descriptions of optical imaging are given in standard books [e. g. 14]. There the image formation is given in terms of an integral of a field amplitude over the lens plane, subject to the assumption that the limiting aperture stop for the system is located at the lens plane. Entangled ghost imaging was analyzed analytically in [16]. There an infinite plane wave pump was assumed and the exit face of the SPDC crystal was integrated over to leave an integral over the lens plane in a manner similar to that of conventional treatments of optical imaging. Here we envision that the effective aperture stop is imposed by the diameter of the pump beam at the SPDC crystal and not by the lens diameter. Therefore we integrate over the lens plane to illustrate the imaging properties and effective limitations on image resolution of the system treated in this paper.

The idler detector operator at the slit plane can be represented in terms of the idler operator at the exit face of the SPDC crystal as

$$E_i^+(x_{1i}, z_{1i}) = \int dx_{oi} h(x_{1i}, z_{1i}, x_{oi}, z_o) a_i(x_{oi}, z_o) \tag{B1}$$

where

$$h(x_{1i}, z_{1i}, x_{oi}, z_o) = \int dx_L \, e^{i\pi(x_L - x_{1i})^2/\lambda_i S_o} e^{i\pi(x_L - x_{oi})^2/\lambda_i d_1} e^{-i\pi x_L^2/\lambda_i f} \tag{B2}$$

and $x_L$ is the transverse coordinate in the plane of the imaging lens.

The integration is accomplished by writing the phase of the integrand in (B2), $\phi$, as a function of $x_L$, identifying the value $x_{L0}$ for which $\dfrac{\partial \phi}{\partial x_L} = 0$ and setting

$$\int dx_L e^{i\phi(x_L)} = e^{i\phi(x_{L0})} \tag{B3}$$

For the function in B2, $\phi(x_L)$ is given by

$$\phi(x_L) = \left(\frac{\pi x_L^2}{\lambda_i}\right)\left(\frac{1}{S_0} + \frac{1}{d_1} - \frac{1}{f}\right) - \left(\frac{2\pi x_L}{\lambda_i}\right)\left(\frac{x_{1i}}{S_0} + \frac{x_{oi}}{d_1}\right) + \frac{\pi x_{1i}^2}{\lambda_i S_0} + \frac{\pi x_{oi}^2}{\lambda_i d_1} \tag{B4}$$

If we use the thin lens relation from equation 1 we find the position of the phase extremum

$$x_{LO} = \frac{d_1(d_1 + d_2)}{d_2}\left(\frac{x_{1i}}{S_0} + \frac{x_{oi}}{d_1}\right) \tag{B5}$$

The phase at the extremum in (B5) is

$$\phi(x_{LO}) = \left(\frac{\pi x_{1i}^2}{\lambda_i S_0}\right)\left(\frac{d_2 - M d_1}{d_2}\right) - \frac{\pi x_{oi}^2}{\lambda_i d_2} - \frac{2\pi M x_{1i} x_{oi}}{\lambda_i d_2} \tag{B6}$$

where M is the magnification given in equation 2. Equations B6, B3 and B2 were used along with equation 5 to derive equation 14 in the main text.

# APPENDIX C: INTEGRATION OVER THE COLLECTING LENS BY STATIONARY PHASE.

The procedure in Appendix C is similar to that used in Appendix B. The operator for the idler detector in the focal plane of the collecting lens is given by

$$E_i^+(x_{D1}, f_c) = \int dx_{1i} h(x_{D1}, f_c, x_{1i}, z_{1i}) E_i^+(x_{1i}, z_{1i}) \tag{C1}$$

The impulse response $h(x_{D1}, f_c, x_{1i}, z_{1i})$ is given by

$$h(x_{D1}, f_c, x_{1i}, z_{1i}) = \int dx_c e^{i\pi(x_{D1}-x_c)^2/\lambda_i f_c} e^{i\pi(x_c - x_{1i})^2/\lambda_i f_c} e^{-i\pi x_c^2/\lambda_i f_c} \tag{C2}$$

where $x_c$ is the transverse coordinate in the plane of the collector lens. The phase in equation C2 is given by

$$\phi(x_c) = \left(\frac{\pi}{\lambda_i f_c}\right)(x_{1i}^2 + x_c^2 + x_{D1}^2 - 2x_c x_{1i} - 2x_c x_{D1}) \tag{C3}$$

The value of $x_c$ where $d\phi/dx_c = 0$ is

$$x_{co} = x_{1i} + x_{D1} \tag{C4}$$

The phase at the extremum in equation C4 is

$$\phi(x_{co}) = -\frac{2\pi x_{1i} x_{D1}}{\lambda_i f_c} \tag{C5}$$

Equations C5 and C2 were used to derive equation 15 in the main text.